\documentstyle[12pt]{article}

\def\beginapjbib{\begingroup \section*{\large \bf References}
         \parskip=.5ex plus 1.0pt
         \def\bibitem{\par \noindent \hangindent\parindent
                \hangafter=1}}
\def\endapjbib{\par \endgroup}

\def\tee{\hbox{$\cal T$}}
\def\oav{\langle O/H \rangle}

\def\li#1{\hbox{${}^{#1}$Li}}
\def\be#1{\hbox{${}^{#1}$Be}}
\def\b#1{\hbox{${}^{#1}$B}}
\def\beq{\begin{equation}}
\def\eeq{\end{equation}}
\def\beqar{\begin{eqnarray}}
\def\eeqar{\end{eqnarray}}
\def\dlibe{\hbox{$d \/{\rm Li}/d\/$Be}}
\def\dbbe{\hbox{$d {\rm B}/d {\rm Be }$}}
\def\grams{\hbox{$\rm \, g \, cm^{-2}$}}

\def\la{\mathrel{\mathpalette\fun <}}
\def\ga{\mathrel{\mathpalette\fun >}}
\def\fun#1#2{\lower3.6pt\vbox{\baselineskip0pt\lineskip.9pt
  \ialign{$\mathsurround=0pt#1\hfil##\hfil$\crcr#2\crcr\sim\crcr}}}

\begin{document}
\thispagestyle{empty}

\rightline{FERMILAB-Pub-94/010-A}
\rightline{astro-ph/9405025}
\rightline{to appear in the Nov.~1, 1994 issue of {\it Astroph.~J.}}

\vspace{.2in}

\begin{center}

{\bf COSMIC RAY MODELS FOR EARLY GALACTIC LITHIUM, BERYLLIUM, AND
BORON PRODUCTION}\\

\vspace{.2in}

Brian D. Fields,$^{1}$ Keith A. Olive,$^{2}$ and
David N. Schramm$^{1,3}$

\vspace{.15in}

{\it
$^{1}$
The University of Chicago, Chicago, IL, 60637-1433 \\
$^{2}$
School of Physics and Astronomy,
The University of Minnesota,
Minneapolis, MN  55455 \\
$^{3}$
NASA/Fermilab Astrophysics Center,
Fermi National Accelerator Laboratory,
Batavia, IL  60510--0500}

\end{center}

\vspace{.2in}

\centerline{\bf ABSTRACT}
\bigskip

\noindent To understand better the early galactic production of
Li, Be, and B by cosmic ray spallation and fusion reactions,
the dependence of these production rates on cosmic ray
models and model parameters is examined.
The sensitivity
of elemental and isotopic production to the
cosmic ray pathlength magnitude and energy dependence,
source spectrum,
spallation kinematics, and cross section
uncertainties is studied.
Changes in these model features, particularly those features
related to confinement, are shown to alter the Be- and B-versus-Fe
slopes from a na\"{\i}ve quadratic relation.
The implications of our results for the diffuse $\gamma$-ray
background are examined, and
the role of chemical evolution
and its relation to our results is noted.
It is also noted that
the unmeasured high energy behavior of $\alpha+\alpha$ fusion
can lead to effects as large as a factor of 2 in the resultant yields.
Future data should enable Population II Li, Be, and B abundances to
constrain cosmic ray models for the early Galaxy.

\bigskip

\noindent {\it Subject Headings} :
Cosmic rays---Galaxy:evolution---nuclear
reactions, nucleosynthesis, abundances

\newpage
\pagestyle{plain}
\newpage
\section{Introduction}

The Population I abundances of \li6, Be, and B
(LiBeB) have been thought for some time to have their origin in
spallation and fusion processes between cosmic ray and interstellar
medium (ISM) nuclei
(see, e.g., Reeves, Folwer, \& Hoyle 1970;
Meneguzzi, Audouze, \& Reeves 1971;
Walker, Mathews, \& Viola 1985).
In the past few years these elements have been sought
in extreme Population II
dwarfs, stars well known to exhibit the ``Spite plateau''
in lithium (Spite \& Spite 1982),
which is understood to indicate the primordial
\li7 abundance (Walker, Steigman, Schramm, Olive, \& Kang 1991).
Recently these same Pop II stars
have also been shown to
contain beryllium,
(Rebolo et al.\ 1988a; Ryan et al.\ 1990, 1992;
Gilmore et al.\ 1992a, 1992b; Boesgaard \& King 1993)
boron, (Duncan, Lambert, \& Lemke 1992)
and most recently \li6 (Smith, Lambert, \& Nissen 1992).  These abundances
provide important clues about the early galaxy, and
cosmic rays have been considered the most likely
production mechanism for LiBeB in these stars as well
(Steigman \& Walker 1992 (SW);
Prantzos, Cass\'{e}, \& Vangioni-Flam 1992 (PCV);
Walker et al.\ 1993 (WSSOF);
Steigman et al.\ 1993 (SFOSW)).

If indeed the extreme population II LiBeB arise from cosmic ray
interactions,
the study of their isotopic abundances opens an
important window on astrophysics.  In principle, we
may be able to gain insight on early cosmic rays,
as well as early star formation rates and chemical evolution
(e.g.\ PCV; Fields, Schramm, \& Truran 1993; Silk \& Schramm 1992).
In addition, associated with these early cosmic ray events is an
appreciable gamma-ray flux
which would contribute (perhaps significantly)
to the present diffuse gamma-ray
background (Silk \& Schramm 1992; Prantzos
\& Cass\'{e} 1993).
The early cosmic ray scenario is also of
importance to cosmology.    One may use the
results of a galactic cosmic-ray (GCR)
spallation and fusion model
to help infer primordial \li7 from the observed Li abundance
(Olive \& Schramm 1993).
One may also use the GCR model's
cosmic ray spallation information on the Li isotopes, compared
with their observed abundances, to
deduce the amount of possible stellar depletion (SFOSW).

Our previous work (WS, WSSOF, SFOSW) was an attempt at a relatively
model-independent approach.
Without assuming a specific model
(of cosmic ray or galactic chemical evolution)
we concentrated on testing the consistency of standard GCR models with
the recent LiBeB observations and with primordial nucleosynthesis
where possible by examining elemental and isotopic abundance ratios.
In another approach,
Gilmore et al.\ (1992b), Feltzing \& Gustaffson (1994),
Prantzos (1993), Prantzos \& Cass\'e (1993),
Pagel (1993), and particularly
PCV have chosen to adopt
a particular detailed model of early galactic chemical and cosmic ray
evolution to examine its predictions and compare with observations.

In this paper we will examine
in detail the dependence of the LiBeB abundances
produced by GCR nucleosynthesis
on the uncertainties of, or allowed variations in,
the cosmic-ray model. As such, we will
discuss the uncertainties in
cosmic ray models.
After a brief review of the data, we present some model
options in section \ref{sect:cosmod}.
Our assumptions regarding the evolution of the abundances
of $\alpha$, C, N, and O
are discussed in section \ref{sect:chemev}.
We address the issue of the
Be and B slopes versus [Fe/H] in section
\ref{sect:slope}.
The cosmic-ray spectrum and
the confinement of cosmic rays will be
discussed in section
\ref{sect:confspec}.
We also explore implications of
these models on $\gamma$-ray production
in section \ref{sect:gammaray}.
We draw
conclusions in section \ref{sect:con}.

\section{LiBeB Abundance Data}
\label{sect:data}

To make this work as self-contained as possible,
we show in table I
LiBeB isotopic abundances observed
in Pop II halo dwarf stars.
We list those stars in which at least two light elements
have been observed as we will
for the most part be primarily interested in elemental or, even better,
isotopic ratios. The notation
we employ below and use throughout the paper is that [X/H]
represents the log abundance
relative to the solar abundance, namely
$\log({\rm X/H}) - \log({\rm X/H})_\odot$ and
[X] $=12 + \log$(X/H). In the table, the iron abundance
represents an unweighted ``world''
average.  For the other abundances, a weighted average is given.
The \li6 abundance was taken from
Smith et al.\ (1993). The \li7 abundances were
taken from Spite \& Spite (1982,1986); Spite et al.\ (1984);
Hobbs and Duncan (1987);
Rebolo et al.\ (1988b); Hobbs \& Thorburn (1991); and
Pilachowski et al.\ (1993).
The \be9 abundances were taken from  Rebolo et al.\ (1988a);
Ryan et al.\ (1992); Gilmore et al.\ (1992a,b);
Molaro, Castelli, \& Pasquini
(1993); and Boesgaard \& King (1993).
Finally, the boron abundances were taken from Duncan et al.\ (1992).

The ratios of \li6 to \li7, Li to Be and of B to Be are the observed
ratios. Because, for Pop II,
the dominant contribution to the \li7 abundance comes from
primordial nucleosynthesis
rather than GCR nucleosynthesis, a certain degree
of caution is necessary when
comparing the first two of these ratios to the
predictions we discuss below. On the other hand
because there is no appreciable
big bang source for either Be or B
(Thomas et al.\ 1993), the ratio of these
two may be compared (unless there is an additional primary source for
\b{11} as discussed
in Dearborn et al.\ (1988), Woosley et al.\ (1990),
and Olive et al.~(1993)).

\footnotesize
\begin{table}[htb]
\begin{center} {\footnotesize \sc Table 1.  Observed Pop II
abundances of LiBeB isotopes}
\end{center}
\begin{center}
\begin{tabular}{lccccccc}
\hline \hline
{\sc Star}& [Fe/H] & Li & [Be] & [B] &
\li6/\li7 & Li/Be${}^{*}$  & B/Be${}^{*}$  \\
\hline
HD16031 & -1.9 & $2.03 \pm 0.2$  & $-0.37 \pm 0.25$ &
        & & $251 \pm 185$ & \\
HD19445 & -2.1 & $2.07 \pm 0.07$ & $-0.14 \pm 0.1 $ &
 $0.4 \pm 0.2$  & & $162 \pm 46$ & $3.5 \pm 1.8$ \\
HD84937 & -2.2 & $2.11 \pm 0.07$ &
  $ -0.85 \pm 0.19$ & & $0.05 \pm 0.02$ & $912 \pm 425$ \\
HD94028 & -1.6&$2.10\pm0.08$&$ 0.44\pm0.1$& & &$46\pm13$ \\
HD132475 & -1.6 & $2.05 \pm 0.09$ &  $0.60 \pm 0.3$  &
        & & $28\pm20$\\
HD134169 & -1.2 & $2.20 \pm 0.08$ &  $0.71 \pm 0.11$  &
        & & $31\pm10$\\
HD140283 & -2.6 & $2.08 \pm 0.06$ & $-0.87 \pm 0.11$
 & $-0.1 \pm 0.2$& &$891\pm262$&$6\pm3$\\
HD160617 & -1.9 & $2.22 \pm 0.12$ & $-0.47 \pm 0.18$ &
        & &$490\pm244$\\
HD189558 & -1.3 & $2.01 \pm 0.12$ &  $0.89 \pm 0.33$  &
        & &$13\pm11$\\
HD194598 & -1.4 & $2.00 \pm 0.2$ & $0.37 \pm 0.12$ & & & $43\pm23$\\
HD201891 & -1.3 & $1.97 \pm 0.07$ &  $0.65 \pm 0.1$
  & $1.7 \pm 0.4$ & & $21\pm6$ & $11\pm11$\\
BD23$^\circ$3912 & -1.5 & $2.37 \pm 0.08$ & $0.30 \pm 0.4$  &
        & &$117\pm110$\\
\hline
\end{tabular}
\end{center}
${}^{\rm *}$Ratios are extremely uncertain due to
inadequate treatment of systematics. \\
\hphantom{${}^{\rm *}$}Formal errors on ratios are underestimates.
\end{table}
\normalsize

Because of the importance of the B/Be ratio, we note that the values
given in  the table for Be and the ratio B/Be, are averages over
observations by several groups.  These observations themselves show
some spread which may be significant.  For the star HD19445,
beryllium upper limits were obtained by Rebolo et al.~(1988a),
giving [Be] $< 0.3$, and B/Be $> 1.3$, Ryan et al.~(1990)
found the upper limit [Be] $< -0.3$ and hence B/Be $> 5$ which should be
compared with the value of 3.5 in the table which represents the only
positive identification of Be in this star by Boesgaard and King
(1993) and is slightly discrepant with the upper limit of Ryan et al.
For HD140283, we have measurements of Be by three groups:
[Be] $= -1.25 \pm 0.4$ from Ryan et al.~(1992) giving B/Be $= 14 \pm 14$;
[Be] $= -0.97 \pm 0.25$ from Gilmore et al.~(1992) giving B/Be $= 7 \pm 5$;
[Be] $< -0.90$ from Molaro, Castelli, \& Pasquini (1993), giving
B/Be $> 6$;
and [Be] $= -0.78 \pm 0.14$ from Boesgaard and King (1993) giving
B/Be $= 5 \pm 3$. Finally for HD201891, [Be] $= 0.4 \pm 0.4$ from
Rebolo et al.~(1988a)  giving B/Be $= 20 \pm 26$ and
[Be] $= 0.67 \pm 0.1$
{}from Boesgaard and King (1993) giving B/Be $= 11 \pm 10$.
As one can see there appears to be a wide range in values
(and uncertainties) in this ratio,
and the values given my be
completely dominated by systematics which are poorly accounted
for in the stated error.\footnote{Since originally submitting this
paper we have learned the results (Kiselman 1994)
of a non- local thermodynamic equilibrium (NLTE)
calculation of the boron abundance in HD 140283.  This calculation
suggests that the boron abundance calculated in the LTE approximation
might {\it underestimate} the actual boron content in the star by
by factors of 3-4.  Because this result relies on delicate atomic modelling,
it should be regarded with caution; nevertheless, if this is the first
indication of a high Pop II B/Be ratio, it can have
intriguing implications for the sources of boron (see Fields, Olive,
\& Schramm 1994).  In addition, the re-analysis
and also points up the potential uncertainty in these difficult measurements.}
In particular, ratios are most useful
if constant surface temperature and surface gravity assumptions
are used in the specific elemental determinations.  Unfortunately,
this is not yet the case.

Indeed it is the uncertainties in these measurements
and in the averages of these
measurements which are themselves uncertain.
Namely, the {\em treatment} of systematic
errors is far from systematic. The conversion of line strengthes to abundances
involves inferences on the surface temperature and surface
gravity of the star.  Most observational determinations have
been made using different sets of inputs.
Though one can ascribe some uncertainty to chosen values of these inputs,
it is not always clear to what extent these systematic
errors have been incorporated into the quoted so-called statistical error,
and different authors make divergent assumptions on the uncertainty of
their assumed stellar parameters.
Furthermore any average will surely underestimate the true error
because of the mistreatment of systematic errors.
Thus it is our feeling that B/Be and Li/Be ratios are extremely
uncertain.

\section{Cosmic Ray Models for LiBeB Spallation Production}
\label{sect:cosmod}

Currently studied models for LiBeB production are based on the work
of Reeves, Fowler, and Hoyle (1970) and the more detailed follow-up work
of Meneguzzi, Audouze, and Reeves (1971).
They describe cosmic ray propagation
via the simple leaky box model.  This assumes a spatially homogeneous
distribution of sources, cosmic rays, and interstellar material.
Within this model, the propagation
equation is
\beq
\label{eq:prop}
\frac {\partial N_A}{\partial t} =
J_A
+ \frac {\partial \left(b_A N_A\right)}{\partial T}
- \frac {1}{\tau_{eff}} N_A
\eeq
Here $N_A = N_A(t,T)$ is
the number density of cosmic ray
isotopes $A = LiBeB$ at time $t$
with energy (per nucleon) between $T$ and
$T + dT$.  Note that eq.\ (\ref{eq:prop}) is the general propagation
expression for $A$, a primary or secondary species, with different
terms being important for each species type.
The first term
on the righthand side, $J_A$, includes all sources for $A$:
\beq
\label{eq:sourceterm}
J_A(T)= Q_A(T) +
\sum_{ij} n_j \int_{T_{th}^{nuc}}^\infty dT^\prime \phi_i(T^\prime)
  \frac {d \sigma_{ij}^A}{d T}\left(T,T^\prime\right)  \, \, .
\eeq
Here $Q_A(T)$ is a possible galactic source of $A$, given as a
number rate per volume and per unit energy; and
spallation production of $A$ appears in the
sum which runs over projectiles $i$ and targets $j$.
Additionally, $\phi_i = N_i v_i$ is the cosmic
ray flux spectrum of species $i$ and,
$\sigma_{ij}^A$ is the cross section for
the process $i+j \rightarrow A + \cdots$.
The second term in the propagation equation (\ref{eq:prop})
accounts for energy losses to the ISM, with
$b_A = - (\partial T/\partial t)_A$ allows for ionization
{}from the Galaxy.
The third term of the propagation equation
accounts for catastrophic cosmic ray losses,
\beq
\frac {1}{\tau_{eff}} =
\sum_{i} \sigma_{iA}^{inel} v_A n_i  +
\frac {1}{\tau_{esc}}
\eeq
with $\sigma_{iA}^{inel}$ encoding spallation
losses of $A$ in the ISM, and $\tau_{esc}(T)$
being the lifetime for cosmic rays against escape.

The propagation equation (\ref{eq:prop}) is solved for the case of a
steady state,
$\partial N/\partial t  = 0$, in which cosmic ray
production is in equilibrium with the losses.
One thus obtains
the {\it spectrum} $N$ of these elements, propagated from their source $J$.
At present, we will ignore losses due to inelastic nuclear collisions,
valid when  $\tau_{esc}^{-1} \gg n_{i} \sigma_{iA}^{inel} v_{A} $
(but see section \ref{sect:conf}).
We assume that the primary cosmic ray species p, $\alpha$, and CNO,
have some homogeneous galactic source $J = Q$,
and negligible spallation production or losses:
$\sigma = 0$.
Writing the solution for these
in terms of the cosmic ray flux $\phi = Nv$, we have
\beq
\label{eq:propflux}
\phi_i(T) =
\frac {1}{w_i(T)} \int_T^{\infty} dT^\prime
  q_i(T^\prime) \exp
    \left( - \left[R_i(T^\prime)-R_i(T)\right]/\Lambda \right)
\eeq
where $w_i = b_i/\rho_{ISM}v$, and $q_i = Q_i/\rho_{ISM}$, and
$i = p$, $\alpha$, CNO.
Also, $R_A(T_A) = \int_0^{T_A} dT^\prime/
(\partial T^\prime/\partial X)_{ISM}$
is the ionization range which characterizes the average
amount of material a particle
with energy $T_A$ can travel before ionization losses will stop it
(expressed in \grams, as $X = \rho_{ISM} v t$).
The ionization ranges were taken from Northcliffe and Schilling (1970) for
low LiBeB energies ($< 12$ MeV/nucleon) at which partial
charge screening effects are important, and from Janni (1982) for
higher energies, using the scaling law
\beq
\label{eq:irscaling}
R_A(Z;T) = A/Z^2 \; R_p(T) \,\,.
\eeq
Here $T$ is the kinetic energy per nucleon and $R_p$ is the
proton range.

In figure \ref{fig:fluxnorm} we plot the proton and
$\alpha$ fluxes $\phi_i$
calculated from eq.\ (\ref{eq:propflux}).
To show the effect of energy losses
on the propagated flux, we also plot $\Lambda q_i$, the solution of
eq.\ (\ref{eq:prop}) for negligible energy losses ($b_i \approx 0$).
As we will discuss below in sections \ref{sect:slope} and \ref{sect:conf},
these losses are important at low energies and negligible at high ones.
This is manifest in figure \ref{fig:fluxnorm}, which shows the
scaling $\phi = \Lambda q$ to be followed closely at high energies,
while at low energies the ionization energy losses significantly reduce
the propagated flux from this scaling.  This behavior is qualitatively
similar for the two source types we consider; propagation differences
at low energies
are discussed in section \ref{sect:conf}.

In solving the propagation equation for the secondary elements LiBeB,
the source term $J$ is assumed to have no primary source component,
i.e.\ $Q = 0$.
The exclusive production of these elements occurs via
spallation between the primaries
and the ISM:  $\sigma_{ij}^{A} \ne 0$.
One can write an expression like eq.\ (\ref{eq:propflux}) for LiBeB,
giving their cosmic ray spectrum.
What we wish to know, however, is not this
steady-state spectrum but instead
the amount of LiBeB thermalized and added to the ISM.
To compute this from the
LiBeB spectrum one assumes that all such nuclei below
some threshold kinetic energy $T_{therm}$
are thermalized and then one examines
the LiBeB current $b_A N_A (T_{therm})$
below this threshold.
Below the lowest spallation threshold
there is no source term in eq.\ (\ref{eq:prop}) for LiBeB,
and the ionization loss term is much larger than the escape term.
Thus to a good approximation the propagation equation (\ref{eq:prop})
reads
\beq
\frac {\partial}{\partial T} b_A N_A (T) = 0 \,\,, T \le T_{th}^{nuc}
\eeq
and so the LiBeB current
$b_A N_A (T_{therm})$ is constant for $T_{therm}$ below spallation
production thresholds.
One may thus choose any
$T_{therm} \le T_{th}^{nuc}$ at which to evaluate this current;
we choose ours to be right at threshold.

With this method of computing the production
rate of LiBeB via evaluation of
the subthreshold LiBeB current,
we can write the
rate of LiBeB accumulation in the ISM as
\begin{equation}
\label{eq:Lrate}
\frac{d \, y_A (t)}{dt} =
\sum_{ij} \; y_j(t)
\int_{T_{th}^{nuc}}^\infty \;
    dT \; \phi_i(T,t) \; \sigma_{ij}^A(T) \; S_A \left[ T_A(T),t \right]
\end{equation}
where $A = \li{6,7}$, \be9, $^{10}$B,
$y_A = n_A/n_H$, the $\phi_i$ come from eq.\ (\ref{eq:propflux}),
and we have ignored the small time variation
of $n_H$.

The factor $S_A(T_A,t)$
accounts for the energy loss of nucleus A in the ISM and gives
the probability of its capture and thermalization; it
is a function of the lab energy $T_A$
of the daughter nucleus A, as well as the epoch $t$, and is
given by
\begin{equation}
\label{eq:sfac}
S_A(T_A,t) = \exp \left[
-\left\{ R_A(T_A)-R_A(T_{therm}) \right\} / \Lambda(t) \right]
\end{equation}
These energy
losses in the ISM
serve to trap the spallation products and
so to allow them to contribute to the
LiBeB finally thermalized in the ISM.  This process competes with
the the spallation products' escape from the galaxy, quantified
by the escape rate $\tau_{esc}^{-1}$ of eq.\ (\ref{eq:prop}).
This quantity appears in eq.\ (\ref{eq:sfac}) through
$\Lambda = \rho_{ISM} v \tau_{esc}$ the (possibly)
energy-dependent average pathlength{\footnote {By writing
the argument of the exponential in
$S_A$ as we have in eq.\ (\ref{eq:sfac}), we are assuming $\Lambda$
to be constant in energy.
For further discussion see section \ref{sect:conf}.}},
also in \grams.
Note that as $\tau_{esc}$ sets the scale for the cosmic ray
residence time before escape, $\Lambda$ is the amount of matter
traversed before escape.
As any $T_{therm} \le T_{th}^{nuc}$ gives
$R(T_{therm}) \ll \Lambda$, we may put
$S_A \simeq \exp(-R_A/\Lambda)$.  For ``forward'' kinematics,
with a light cosmic ray nucleus impinging on a stationary,
heavy ISM nucleus, $T_A$ is small for typical spallation
energies, and so $R_A \ll \Lambda$ and
$S_A \simeq 1$.  For ``inverse'' kinematics, with  a heavy cosmic
ray nucleus on a light ISM particle, $S_A$ can differ significantly
{}from unity and the LiBeB yields are reduced accordingly.

Following Walker et al.\ (1992), we put
\begin{equation}
\label{eq:crscale}
q_i(T,t) = y^{CR}_i(t) \; q_p(T,t)
\end{equation}
with $i = {\rm p,} \alpha$, CNO:
i.e., we posit the constancy of the
cosmic ray isotopic and elemental ratios at the source
and over the entire energy spectrum.
We then choose to make the more serious
assumption that we may
express the proton source strength in the separable form
\begin{equation}
q_p(T,t) = q_p(T) \; f(t).
\end{equation}
We will in fact consider various
Population II source spectra $q$ (see section \ref{sect:spec}), but we will
not allow a fully general energy or time dependence.
While this has the immediate advantage of simplifying
the calculations, one may view this assumption
as a postulation that the present mechanism of
cosmic ray acceleration does not differ dramatically
over time in its energetics, but only  in its net
cosmic ray output.

In our analysis of cosmic ray model features, and in the
accompanying figures, we will concern ourselves
with the effect on the LiBeB production rates---or rather
their ratios---as calculated by eq.\ (\ref{eq:Lrate}).
Given a set of cosmic ray and ISM abundances, and a
confinement parameter $\Lambda$, these rates
may be evaluated numerically.
As in previous work, we used empirical values
of partial cross sections tabulated in by Read and Viola (1984).
To proceed further and integrate the rates to get the LiBeB
yields would require a full model for cosmic ray and
chemical evolution.  We discuss this issue in the next section.

As a cosmic ray model of propagation, the above amounts to the
simple leaky box
(reviewed by, e.g., Ceasarsky 1980, 1987).
This is the simplest model used to describe cosmic
rays today, and as such has obvious computational advantages.
While the simplicity of the leaky box makes it a useful tool,
it is physically unrealistic and thus, in some cases,
quite inaccurate.
A proper treatment of
cosmic ray propagation must explicitly include diffusion effects
only sketched with the leaky box confinement parameter $\Lambda$.
Also, a proper model must abandon the simple leaky box assumption
of spatial homogeneity of sources and interstellar material.
Despite these shortcomings, however,
we will follow previous authors and
adopt this model in what follows, both for the above pragmatic
reasons and moreover because the epoch we consider is so poorly understood
in its relevant details that adoption of a more detailed description
is not warranted at the present time.

\section{The Role of Chemical Evolution}
\label{sect:chemev}

Although we can calculate the
rates in eq.\ (\ref{eq:Lrate}), for a given set of parameters,
using well-understood physics, to integrate these rates
requires knowledge of the chemical and dynamical
evolution of the cosmic ray and
ISM species in the Galaxy's early history.  Such knowledge is sketchy
at present, but some reasonable assumptions may prove fruitful.

If we assume that the H and He abundances do not change much from
their values as set by the big bang, we have
\beqar
\nonumber
H(t) & \approx & H^{BB} \\
y_{He}(t) & \approx & y_{He}^{BB} \; \approx \; 0.08  \;.
\eeqar
If we further assume that the CNO nuclides evolve at the same rate,
so that their ratios remain constant within the early epoch considered
here, we put
\begin{equation}
\label{eq:cnorat}
\frac{y_C(t)}{y_C^{OBS}} =
  \frac{y_N(t)}{y_N^{OBS}} =
  \frac{y_O(t)}{y_O^{OBS}} \;.
\end{equation}
Here we denote with OBS the observed CNO abundances for a given Pop II
Be or B star.
Equation  (\ref{eq:cnorat}) is
assumed to hold for times $t$ less than the galactic age $\tau$ at the
birth of the star.
One should interpret equation \ref{eq:cnorat} with some care,
as approximation of constant C:N:O ratios is not strictly true.  We
expect these elements to have different sources:
O is made in type II supernovae
but C and N are made primarily in intermediate mass stars.
Thus any differences in the
evolution of these sites will result in differences in the abundance
ratios.  Fortunately, large differences between CN and O production
are only important at early times, and the observed Pop II CNO
abundances are consistent with constant C:N:O over the timescales
important here.
One should note, however, that in Pop II stars the O/CNFe ratio,
though roughly constant, does exceed the
solar ratio by a factor of $\sim 3$ .
This comes about because of the predominance at this
epoch of type II supernovae over type I's.  The effect is important
for our purposes, as spallation of O gives a lower B/Be ratio than
spallation of C.

We again simplify by extrapolating the present relation
between abundances of cosmic ray and ISM species,
namely we assume
that the relative enhancement of the cosmic ray
CNO abundances
over interstellar CNO abundances
has remained constant in time.
We put
\beq
\frac{y_i^{CR}(t)}{y_i(t)} =
\left( \frac{y_i^{CR}}{y_i} \right)_{present}
\label{eq:cnoprop}
\eeq
where $i = $CNO and the righthand side denotes the present value.
We will in fact adopt the stronger assumption, made in most works,
that
\beq
\label{eq:creqism}
y^{CR}(t) = y^{ISM}(t)
\eeq
but we note that this assertion is false for the present cosmic rays.
In particular,
Asakimori et al.~(1993), and references therein, report
a H/O depletion in present cosmic rays of about a factor of 2 relative to
the solar system ratio.
The proton-to-helium
ratio in the cosmic rays is energy dependent and varies from
$\sim 0.2$ at the lowest measured energies to $\sim 0.05$ at the
highest energies which interest us here.
Additionally, as shown in Buckeley et al.~(1993),
He/O in the cosmic rays is depleted by
a factor of $\sim 4.6$ relative to its local galactic value.
In this paper we
have normalized to the protons, and thus
adopting the observational results would amount
to reducing $y_{\rm He}$ by 2.3,
and enhancing $y_{\rm CNO}$ by a factor of 2.
We will instead use eq.\ (\ref{eq:creqism}) to allow
comparison with previous results.
Bear in mind, however, that the
cosmic ray abundance scalings are uncertain by
at least a factor of 2.

There are proposed scenarios in which
early cosmic ray abundances do not scale with early ISM abundances.
The two suggestions are constructed to allow for spallation
to occur with CNO abundances at high, Pop I levels,
thus providing a natural explanation for the Be/O constancy.
The first of these posits that supernovae
accelerate their own (metal-enriched) ejecta,
which makes the cosmic ray CNO abundance high.  In the case,
LiBeB production is dominated by the spallation
of these fast, heavy nuclei, rather than by spallation
of fast protons and $\alpha$ particles.
We may model this behavior by putting
\beq
y_O^{CR}(t) = y_O^{supernova} \simeq y_O^{present} \gg y_O^{OBS}
\eeq
The other suggestion is that spallation
occurs predominantly in (temporarily)
metal-enriched regions around supernova
remnants.  In this case the cosmic ray abundances do scale with those of
the ISM, since the target abundaces are always at Pop I levels.
In what follows, however, wherever not explicitly
noted otherwise, eq.\ (\ref{eq:cnoprop}) will be assumed to hold.

If we allow the confinement pathlength
to vary with time, then as we will see
below, we may not further simplify the rate equation (\ref{eq:Lrate}),
or the integral that is its solution.  	If the confinement is taken to be
constant in time---a dubious proposition, as we shall see---we
may now separate the {\it propagated} flux:
\beq
\phi_i(T,t) = y_i^{CR} f(t) \varphi(T) \, \, ,
\eeq
where $y_i^{CR}$ is the cosmic ray abundance of $i$ relative to protons,
as in eq.\ (\ref{eq:crscale}).
This allows one to write the solution to the rate equation
as a sum of terms, each of which has a factor that involves
an integral over time and a factor involving an integral over
energy:
\begin{equation}
\label{eq:ysoln}
y_A(\tau) = \sum_{ij} \; \Delta_{ij} \:
\langle \Phi \sigma_{ij}^A \rangle
\end{equation}
with the ``exposure time'' given by
\begin{equation}
\Delta_{ij}(\tau) =
\int_{0}^\tau \; dt \; y_j(t) \;
             y_i^{CR}(t)\; f(t)
\end{equation}
with $\tau$ is the age of the Galaxy at the birth of the star
in question, and the ``reduced rate'' is
\begin{equation}
\langle \Phi \sigma_{ij}^A \rangle =
\int_{T_{th}^{nuc}}^\infty \;
    dT \; \varphi_i(T) \; \sigma_{ij}^A(T) \; S_A(T_A) \;\;.
\end{equation}
Note that these factors
are different for forward and inverse kinematics,
that is, $\langle \Phi \sigma_{ij}^A \rangle_f
\neq \langle \Phi \sigma_{ij}^A \rangle_r$.
As mentioned above, this stems from the longer range
and greater chance of escape for the fast
$A$--nuclei produced in inverse kinematic collisions.

With the assumptions made thus far in this section---most importantly,
that of the time constancy of $\Lambda$,
the number of independent exposure times may be
reduced to three (SW, WSSOF):
\begin{equation}
\Delta_{ij}
      = \left\{ \begin{array}{ll}
	\Delta_{\alpha,\alpha} & i=j=\alpha \\
	\Delta_{p,O}^{ISM}     & \mbox{forward kinematics} \\
	\Delta_{p,O}^{CR}      & \mbox{inverse kinematics}
	    \end{array}  \right.
\end{equation}
Of these, we will take $\Delta_{\alpha,\alpha}$ as
the fiducial quantity, as it measures
the integrated flux enhancement exclusively
(i.e.\ with no explicit dependence on chemical evolution),
and we define
\beq
\tee \equiv \Delta_{\alpha,\alpha}/y_{He}^2
= \int_{0}^\tau \; dt \; f(t)
\eeq
We then define the quantity $\oav$, which is a measure of
the average abundance of oxygen relative to hydrogen,:
\beqar
\nonumber
\oav_{ISM} & \equiv &
             {\Delta_{p,O}^{ISM}} / {\tee}  \\
             & = & \frac{\int_{0}^\tau \; dt \; y_O^{ISM}(t) f(t)}
	           {\int_{0}^\tau \; dt \; f(t)}
\eeqar
with a similar expression for $\oav_{CR}$.
Note that $\oav$ is to the weighted average of O/H
at time $\tau$.  This average weighted
by the net flux enhancement $f(t)$, and is normalized.

With this notation we have
\begin{eqnarray}
y_{Be,B} & = &
\tee
\; \sum_{ij}\vphantom{\sum}^\prime \,
y_i^{CR} y_j^{ISM} \\
\nonumber
& & \hphantom{\tee \: \left[ \right]}
\; \times \;
\left( \frac{\oav_{ISM}}{y_O^{ISM}}
\langle \Phi \sigma_{ij}^{Be,B} \rangle_f +
\: \frac{\oav_{CR}}{y_O^{CR}}
\: \langle \Phi \sigma_{ij}^{Be,B} \rangle_r \right)
\label{eq:ybe}\\
y_{Li} & = &
\tee \:
\left[ y_{He}^2 \langle \Phi
\sigma_{\alpha \alpha}^{Li} \rangle \right. \\
\nonumber
& & \hphantom{\tee \: \left[ \right]} \left. +
\sum_{ij}\vphantom{\sum}^\prime \, y_i^{CR} y_j^{ISM}
\left( \frac{\oav_{ISM}}{y_O^{ISM}} \;
\langle \Phi \sigma_{ij}^{Li} \rangle_f
+ \frac{\oav_{CR}}{y_O^{CR}} \;
\langle \Phi \sigma_{ij}^{Li} \rangle_r \right) \right]
\label{eq:yli}
\end{eqnarray}
where the prime on the summation indicates that any $i=j=\alpha$ term
has been deleted.  Note that the ratio of LiBeB production rates,
i.e.\ of eqs.\ (\ref{eq:ybe})
and (\ref{eq:yli}), does not depend explicitly on $\Delta_{\alpha \alpha}$.
However, the flux enhancement appears via the strong dependence
on $\oav$, which includes a factor
of $\Delta_{\alpha \alpha}$ in its definition.

{}From eq.\ (\ref{eq:ysoln}) it is clear that chemical evolution effects
are encoded into the LiBeB abundances in terms of exposure time, in which
these effects appear in
convolution with cosmic ray flux evolution.  While this means that
the absolute abundances are dependent on chemical
evolution, it also means that the relative abundances are much less so if
the LiBeB isotopes compared originate in similar processes.
This is the case with Be and B, which are both produced in
pure spallation events, as opposed to Li, which is produced not
only by spallation but also in $\alpha + \alpha$ fusion events.
This point was stressed by WSSOF, who noted that the ratio
B/Be is the {\it least} dependent on our knowledge of the early galaxy
and of early cosmic rays.  WSSOF showed that any model for early cosmic ray
LiBeB production yields a B/Be ratio lying within the range
$7.5 \le {\rm B/Be} \le 17$.  One might compare this
with the average of the observed values
in the most extreme Population II dwarf, HD 140283,
which has B/Be = $6 \pm 3$.
This ratio is the strongest prediction
of the cosmic ray model.  However,
while the theoretical prediction is firm, particularly in setting
a lower bound
to the B/Be ratio from cosmic rays,
the experimental ratio is still difficult to obtain reliably.
As discussed in section \ref{sect:data},  the true
errors in the measured value, when including systematic uncertainties,
are certainly larger than the simple weighted mean we quote here.
Comparisons with the observations are thus not all that
compelling at the present time.

Under the assumption that there is no additional (primary) source for
\b{11}, the B/Be ratio is weakly dependent on chemical
and cosmic ray evolution.
However, it is well known that the predicted value of the
\b{10} to \b{11}
ratio in GCR nucleosynthesis is about 2.5, whereas the observed
ratio in meteorites is just over 4.
Dearborn et al.\ (1988) proposed a possible thermonuclear production
of \b{11} in Type II supernovae, and
Woosley et al.~(1992) had suggested
that neutrino processes
during supernovae may yield a significant abundance of \b{11}.
Both suggestions
are tested in a GCR and chemical evolution model by Olive et al.~(1993),
where it was found
that while the \b{11}/\b{10} ratio could be brought
into agreement with the observation,
the supernova sources added a significant primary source for
\b{11} which upsets
the constancy of the B to Be ratio as a function of metallicity.
Though the data seem to disfavor this primary source which predicts a
B/Be ratio
in excess of 50 at [Fe/H] $< -3$, the scant amount of data at present
can not conclusively eliminate this possibility.

In contrast, the
ratios of Li with Be and B are always strongly dependent upon
chemical and cosmic ray evolution
because of
the $\alpha+\alpha$ channel for production of Li.
This is in fact the
most important process for Pop II cosmic ray synthesis of Li, since
in the metal-poor Pop II environment, O/H can be reduced from its
solar value by two or more orders of magnitude, while He/H is undepleted.
In the language of eq.\ \ref{eq:yli}, $\oav \ll y_{He}^2$, and so the only
important contribution to Li synthesis comes from the $\alpha+\alpha$ term.
In addition, the dominant component of the observed \li7 is primordial,
further complicating
comparison of Pop II LiBeB with GCR nucleosynthesis predictions.

\section{The Be and B {\it versus} Fe Relations}
\label{sect:slope}

Many have suggested (e.g.~Vangioni-Flam et al.~1990)
that since Be and B arise from the interaction of
stellar nucleosynthesis products with cosmic rays,
they should vary quadratically with
elements, e.g.\ O, which form their targets.
(Recall that while O/CNFe is enchanced in Pop II
compared to Pop I, the ratios are still roughly constant
and so a relation between LiBeB and CNO
also manifests itself as the same relation between LiBeB and the
more commonly observed Fe.)
This quadratic dependence is argued to follow from the fact that
cosmic ray sources are presumed to be supernovae
which are the producers of the oxygen targets.
However,
cosmic rays
have a large confinement time and hence the instantaneous
recycling approximation that yields this quadratic dependence
is quite inappropriate.  Even more significant is the fact that
the data
does not show this dependence and is instead best fit by a Be slope
relative to O that is much closer to 1 than to 2.  This discrepancy,
has received some attention and has even been
interpreted as a failure of
the cosmic ray model.  Again, this conclusion ignores the effects
of cosmic ray confinement, as we now demonstrate.

Let us now go through the logic in greater detail.
The argument for a quadratic scaling of Be and B with O proceeds from
these assumptions:
({\it a}) Type II supernovae are the dominant source for O, so
one expects a relation to $N_{SN}$, the total integrated number of
supernovae, of the form $O \propto N_{SN}$.  ({\it b}) Supernovae are also
likely to be the acceleration site for cosmic rays, so one might
na\"{\i}vely expect
the cosmic ray flux $\phi$ to vary with the current {\it rate} at which
these acceleration sites are created:  $\phi \propto d\, {N_{SN}}/dt$,
if we ignore confinement and assume instantaneous recycling of
the cosmic rays.
Put slightly differently, in the leaky box model we have adopted,
we have assumed an equilibrium to hold between the rate at which
cosmic rays are lost and the rate at which they are created.
In this model one can sensibly
argue that the cosmic ray flux is indeed proportional
to this creation rate.
Then, taking the essential pieces of the rate equation (\ref{eq:Lrate}),
we obtain the na\"{\i}ve result
\beqar
\label{eq:naive}
\nonumber
\hphantom{\Longrightarrow \,\,}  d {\rm Be}/dt & \propto & {\rm O} \phi  \\
\nonumber
 & \propto & N_{SN} d\, {N_{SN}}/dt \\
\Longrightarrow \,\, {\rm Be} & \propto & {\rm O}^2
\eeqar
which gives the advertised quadratic dependence, with an analogous equation
for B evolution.

While assumption ({\it a}) seems solid, assumption ({\it b})
is not because it accounts
for evolution of cosmic ray sources but not for changes in confinement.
Indeed the evolution of $\Lambda$ is the critical feature allowing
one to avoid the
na\"{\i}ve secondary-to-primary ratio for the B- or Be-to-O ratio
of eq.\ \ref{eq:naive}.
For a given source density $q$ of {\it primary} cosmic ray elements
(at some instant $t$), the propagated
flux $\phi$ increases with $\Lambda$, as at larger $\Lambda$
the cosmic rays traverse more matter before escape and
so the steady state between production and escape is reached at a
higher flux.
In the limit of no energy losses,
valid at high energy, the relation is $\phi = \Lambda q$.
With energy losses,
this relation is energy dependent, but for the total flux
PCV find that
it varies roughly as $\phi \propto \Lambda^{1/2}$.
Thus we should alter the cosmic ray flux scaling
to be $\phi(t) \propto \Lambda^{1/2}(t) N_{SN}(t)$,
and it follows that changes in the early
confinement alter the cosmic ray flux, and its
associated LiBeB production.
Specifically, in the place of eq.\ \ref{eq:naive}, we have
the more general relation
\beqar
\label{eq:lamdepflux}
\nonumber
\hphantom{Longrightarrow} d {\rm Be}/dt & \propto
	& \Lambda^n {\rm O} d\, {\rm O}/dt  \\
\Longrightarrow  d {\rm Be}/d\, {\rm O} & \propto & \Lambda^n {\rm O}
\eeqar
with $n \le 1$ the index for the scaling law between cosmic ray flux
and confinement (with PCV finding $n = 1/2$).
If $\Lambda$ can vary with time---and indeed it is hard to imagine how
to avoid such variance---then the simple quadratic relation between
Be or B and O is clearly not the prediction of the cosmic ray model.
The effect of the variance of $\Lambda$ is further discussed in
the next section.

\section{Cosmic Ray Model Dependences}
\label{sect:confspec}

We wish to investigate the dependence
of calculated elemental ratios Li/Be and B/Be
on various model parameters,  for
different epochs, i.e. different [Fe/H].  Such a calculation, however, requires
specific models for galactic chemical and cosmic ray evolution.
Without constructing such models, what
we can examine, at a given [Fe/H], are the ratios
of production {\it rates}
\beqar
\frac {d y_{\rm Li}/dt}{d y_{\rm Be}/dt} \equiv \dlibe\
& {\rm and} &
\frac {d y_{\rm B}/dt}{d y_{\rm Be}/dt} \equiv \dbbe
\eeqar
(c.f.\ eq.~(\ref{eq:Lrate}))
and their dependence on escape length.

This comparison of production rates provides
useful bounds on the elemental ratios any full model
would predict.  Specifically,
at a given epoch [Fe/H] and confinement $\Lambda$,
\dlibe\ gives the value  Li/Be would take if
the ISM abundances and cosmic ray confinement
had been constant for the entire history of the galaxy up to that epoch.
The ISM abundances of heavy elements CNO certainly are not constant but
build up from zero, while the ISM abundances of H and He do remain roughly
constant.  The chemical evolution of the ISM
affects Li and Be or B differently; Li
is largely unaffected, being made mostly by $\alpha+\alpha$ fusion, while
Be and B are made by spallation of the
CNO targets whose abundances will have
always been rising to the levels observed at a particular [Fe/H].
The calculated \dlibe\ will thus provide a {\it lower} bound on the
actual Li/Be a full chemical evolution model would predict
for a given [Fe/H].{\footnote {Just as the
CNO abundances will have changed, so too
the cosmic ray
flux strength and confinement will have evolved.  Though an overall
change in flux strength does not affect elemental ratios,
evolution in confinement will change the spectrum, with high
early confinement leading to a harder spectrum in the past (see next section).
This {\em decreases} the Li/Be ratio, as the $\alpha+\alpha$ production
falls off at high energies.  This goes in the opposite direction as
the chemical evolution effect.  However, we will in general explicitly
examine the allowed confinement parameters, and in this context
we may understand the \dlibe\ ratio to be bounded from below.}}
The B/Be ratio,
however, does not change at all with CNO evolution as long as
the C:N:O ratios remain constant, and thus is equal to \dbbe.
These ratios are roughly constant over the Pop II epoch we consider here,
but they do of course change eventually
as the O overabundance diminishes to it Pop I value.

\subsection{Confinement}
\label{sect:conf}

PCV have detailed a chemical evolution model
that can fit observed Be and B data
and does produce a slope versus iron that is smaller than
the na\"{\i}ve quadratic dependence.  In their model, the pathlength is
assumed to be larger than the present $\sim 10 {\rm g cm}^{-2}$, due to
the larger confinement volume of the halo phase versus the disk phase, as
well as the larger early gas fraction.  They suggested that the confinement
varies over the range
$10 \grams \la \Lambda \la 1000 \grams$.  This variation
arises specifically from scaling the escape length
as $\Lambda \propto \sigma_g \, H$; where
the gas mass fraction $\sigma_g$
decreases with time according to the adopted
chemical evolution, while the disk scale height $H$ has an exponentially
decreasing evolution imposed on it by hand,
to simulate the dynamics of the
halo collapse.

While this line of reasoning is attractive, one must be cautiously aware of
the uncertainties involved.
For example, the early behavior of the main confinement
mechanism, the galactic magnetic field, is not well understood.
The favored means for field generation is through dynamo action
tied to the galactic rotation, setting a timescale for
generation of $\tau_{rot} \sim 10^8$ yr.
Note that in this case the galaxy does not as efficiently retain LiBeB
long enough to thermalize them,
particularly given the flatter (harder) cosmic ray spectra
and/or the inverse kinematics some have suggested (see section
\ref{sect:spec}).

Furthermore, Malaney and Butler (1993) have pointed out that the escape
length is bounded from above.  Physically, this happens when the confinement
becomes so large that nuclear destruction of fast LiBeB becomes important.
In terms of eq.\ (\ref{eq:prop}), we have
\beq
\label{eq:lameff}
\frac{1}{\Lambda} = \frac{1}{\Lambda}_{conf} + \frac{1}{\Lambda_{nuc}}
\eeq
where $\Lambda_{nuc}^{-1} = \sum_i n_i \sigma_{iA}^{inel} / \rho_{ISM}$.
Malaney and Butler note that $\Lambda_{nuc}$ is species-dependent
(whereas $\Lambda_{conf}$ is not), and they
estimate $\Lambda_{nuc} \la 200 \grams$ for protons, thus setting
an upper limit for $\Lambda$.   Prantzos and Cass\'{e} (1993) have included
this effect in the PCV model and find it makes little change in their
results.

What is the effect of a changing confinement?
A variation in confinement acts not only to alter the
flux intensity but also to change the shape of the
spectrum.  This behavior is apparent in figures
\ref{fig:fluxlam}({\it a}) and ({\it b}), which plot spectra
for a fixes source strength $q$ but with
different values of confinement $\Lambda$.  Notice that in both plots
the total flux
is higher for larger confinement, but by different amounts in different
energy regimes.  The lowest energy ($\la 100$ MeV/nucleon)
portion does not enjoy
the same enhancement as the high energy ($\ga 1$ GeV/nucleon) region.
One can see in eq.\ (\ref{eq:propflux})
that the spectrum
turns over roughly at the energy at which $R(T) \sim \Lambda$.
A spectral peak
at a larger $T$ means
increased power at higer energies.
In other words, increasing confinement enhances the highest energy
regime and so hardens the spectrum.

Note the different low-energy flux behaviors for the two source types
(fig.\ \ref{fig:fluxnorm}),
the forms of which are given explicitly below (section \ref{sect:spec},
eqs.\ (\ref{eq:psource}) and (\ref{eq:esource})).
The momentum source (broken curve in fig.\ \ref{fig:fluxnorm}({\it a})),
diverges as $T \rightarrow 0$,
whereas the energy source (broken curve in fig.\ \ref{fig:fluxnorm}({\it b})),
is finite for small $T$.
These differences manifest themselves in the propagated fluxes.
The flux (solid curve in fig.\ \ref{fig:fluxnorm}({\it a})),
derived from the momentum source
has a mild turnover at small energies,
and the different $\Lambda$ fluxes all converge.  This
behavior is contained in the flux equation (\ref{eq:propflux}).  At low energy
($T \ll m_p$), the escape length
$R(T) \sim T/(\partial T/\partial X) \ll \Lambda$, is slowly varying,
and so the exponential in eq.\ (\ref{eq:propflux}) changes slowly as well.
On the other hand, the source is a featureless and
steeply falling
power law: $q(p) \sim p^{-2} \sim T^{-1}$.
Thus one can, at sufficiently low energy,
ignore the confinement parameter altogether and obtain
\beq
\label{eq:philowe-psor}
\phi(T) \approx \frac{1}{w(T)} \int_T^\infty dT^\prime \, q(T^\prime)
\eeq
which is independent of confinement.
For a source power law in total energy,
however, the source
has a characteristic scale $m_p$ and
does not strongly vary over this scale.
Since $R(T)/\Lambda$ does change appreciably in this range,
eq.~(\ref{eq:philowe-psor}) is not appropriate, and one must use the full
flux equation, with its the strong dependence on confinement.
Thus the propagated fluxes (solid curve in fig.\ \ref{fig:fluxnorm}({\it b})),
do not display the same convergence as those
derived from a momentum-law source.

The well-known (e.g.\ PCV)
dependence of the propagated flux on confinement
has important consequences for elemental
ratios.  Although
changes in the total cosmic ray intensity will not affect
elemental production ratios,
spectral changes can.  One calculates LiBeB production rates by integrating
the product of
cross section with flux over kinetic energy.
Thus the energy regimes in which this product is the largest will
contribute most to the production rates.
As shown if fig. \ref{fig:fluxnorm}, the flux peaks either at
$\sim 100$ MeV/nucleon (for a momentum spectrum), or at
$\sim 1000$ MeV/nucleon (for a total energy spectrum).
This behavior is to be cmopared to that of spallation and fusion
cross sections, which typically
display low energy resonance peaks up to a few hundred MeV/nucleon,
followed by a high energy asymptotic behavior
characteristic of the interaction.
Specifically, the cross sections approach a constant at high energies
for spallation
interactions with a many body final state, but suffer an exponential dropoff
for fusion reactions which have few-body final states.
(It should be cautioned here that the continued exponential falloff
of the $\alpha+\alpha$ fusion reactions above a few hundred MeV
is only an extrapolation since no data exists at higher energies;
c.f.\ section \ref{sect:omod}.)
Thus a change in confinement has the effect on
\dbbe---the ratio of pure spallation products---of
changing the emphasis on the resonance region
relative to the plateau.
Both $p$ and $\alpha$ cross sections for B production have lower thresholds
than those for Be, and B cross sections also display higher peaks relative
to their plateaus.
Consequently, the \dbbe\ ratio is lowest at cosmic ray
energies in the plateau region.  To produce a low B/Be ratio requires
a spectrum that gives a larger weight to this energy regime, i.e.\
a harder spectrum.
Just such a spectrum arises from a large confinement scale.
The absolute effect on
the \dbbe\ is small, but important given that B/Be is constrained
theoretically and observationally to lie in a small range.

For the \dlibe, B production ratio,
spectral flattening via large confinement
leads to more pronounced changes.
The effect again stems from the
high energy cross section behavior:  while the Be and B cross sections are
asymptotically flat, the $\alpha+\alpha \rightarrow Li$ fusion
cross sections (which dominate the Pop II cosmic ray production of Li)
are assumed to
fall off exponentially at high energy.  Thus harder spectra have a strong
effect \dlibe\ as they emphasize an energy regime where the Li production
is not just reduced relative to Be, but in fact negligible.  Thus
\dlibe\ strongly decreases with
increasing confinement.

We have computed numerically the ratios of rates for
the range $10 \grams \le \Lambda \le 1000 \grams$.  The results appear in
figures \ref{fig:lb-lam} and \ref{fig:bb-lam}, and are in agreement with
our expectations.  The strong dependence of these ratios
on the different forms of
source spectra will be further discussed below
(section \ref{sect:spec}).
In reading these figures and comparing to observational data,
one must bear in mind the need to include evolutionary effects properly
in order to go from the plotted ratios of rates
to elemental ratios.
Furthermore, as argued above, we see the falloff of both \dlibe
and \dbbe\ at large $\Lambda$.  Note that the scales in these figures
are very different.  As $\Lambda$ is increased from 10 to 1000\grams,
\dlibe\ falls off by a factor of 10 (fig.~ \ref{fig:lb-lam}),
whereas \dbbe\ drops off by only
about 15\% for the momentum source spectrum and even less so for
the energy source spectrum (fig.~\ref{fig:bb-lam}).

In figure \ref{fig:lb-fe} we plot the production ratio
\dlibe\ against [Fe/H].
Such a plot suggests a comparison with  observational data, but
bearing in mind the caveat that
the \dlibe\ value is only a lower limit to the
actual value a full chemical evolution model would
give.  See the end section \ref{sect:spec} for a discussion of such a
comparison.  Also note that a plot of \dbbe\ as a function of [Fe/H]
is uninteresting as long as one assumes the C:N:O:Fe ratios remain
constant during the Pop II epoch, since in this case the \dbbe\ ratio
remains constant (for a fixed confinement strength).

In our considerations thus far we have followed most previous work on early
LiBeB in putting $\Lambda$ to be a constant in energy.  This parameterization
offers a simple way of trying to characterize the early galactic conditions
which, as has been pointed out, are poorly understood.  While
offering simplicity, this approach does suffer the criticism that
the choice of a constant escape length does not fit the observed
present day cosmic ray pathlength distribution.
As shown most recently in Garcia-Mu\~{n}oz et
al.\ (1987), the escape length varies with energy, increasing slowly to a
maximum around a GeV/nucleon, then decreasing as $T^{-0.6}$.
They find that their empirical form for $\Lambda(T)$ is well fit by
\beq
\label{eq:gmpld}
\Lambda(T) = \Lambda_0
  \left( \frac{T+m_p}{T_1+m_p} \right)^\beta
  \left[ 1 - 0.2\exp \left( -
           \frac {\left| T-T_1 \right|}{T_2} \right) \right]
\eeq
with $\Lambda_0 = 11 \grams$, $T_1 = 850$ MeV, and
$T_2 = 300$ MeV; $\beta = +2$ for $T < T_1$, and $\beta = -0.6$
for $T \ge T_1$.  A plot of this function appears in fig.\ \ref{fig:gmpld}.
As is well-known, an energy-dependent pathlength changes the solution to
the propagation equations (\ref{eq:prop} and \ref{eq:Lrate}),
with the replacements
\beq
\label{eq:lamoft}
R(T)/\Lambda
     \longrightarrow
\int_0^T \frac{dT^\prime}{w(T^\prime) \Lambda(T^\prime)}
\eeq
where $w(T) = (\partial T/\partial t)/\rho_{ISM} v$.

We have allowed the normalization parameter $\Lambda_0$ in $\Lambda(T)$
to vary, and figures \ref{fig:lcomplibe} and \ref{fig:lcompbbe}
compare the resulting  $\dlibe$ and \dbbe\
to the same ratios obtained with a constant $\Lambda(T)=\Lambda_0$.
Note that the changes are relatively small.  This is because the
energy dependence of the pathlength is most important at high
energies, above the region $R(T) \sim \Lambda(T)$ at which the flux
turns over.  However, the majority of the flux resides in the
region around and below this maximum.  Since these energy regimes
are largely unaffected, the LiBeB production is weakly affected as well.
For the momentum
source spectrum the results with and without an energy-dependent
pathlength differ by a constant factor of roughly 25\%, while for the
energy source spectrum the difference varies but is always small $\la 5\% $.
The smallness of the difference between
the two confinement behaviors
justifies the $\Lambda(T) = \Lambda_0$ approximation
used in this work and elsewhere.  The systematic decrease of the yields
for a given scale $\Lambda_0$ follows from the energy dependence, under
which $\Lambda$ has a maximum, at $T=T_1$, of only $0.8 \Lambda_0$.  Thus
the confinement is always less than this value, and
very much lower for the important low energies.

We have investigated the effect of inelastic nuclear reactions between
the primary cosmic rays and the ISM on the effective confinement
given in eq.\ (\ref{eq:lameff}).
As discussed by Malaney and Butler (1993), this effect
contributes an extra
species-dependent and energy dependent term to the escape length.
Because the pathlength is energy dependent, one must use eq.\
(\ref{eq:lamoft}) to compute its effect.  We have done so,
using cross sections given in Meyer (1972), Townsend \& Wilson (1985),
and Bystricky, Lechanoine-Leluc, \& Lehar (1987).
The results appear in Figure \ref{fig:lamwinel}, and the behavior is
as expected.  In the usual case, \dlibe\ decreases with increasing
$\Lambda_{eff} = \Lambda_{esc}$, and we now have
$\Lambda_{eff} < \Lambda_{esc}$.  Thus, we expect
\dlibe\ to be consistently larger at a given $\Lambda_{esc}$, as is
the case.  This effect is more pronounced at high $\Lambda_{esc}$, where
the nuclear loss term dominates.

Another property of the early galaxy that affects confinement is
the state of ionization of the ISM.
Energy losses are increased in an ionized medium,
leading to a shorter ionization range.  Because the confinement
always appears as $R/\Lambda$ (for $\Lambda(T)$ constant in energy), this
effectively increases the escape length
PCV account for ionization by
doubling the energy losses for neutral media, a procedure which amounts
to assuming a fully ionized medium (Ginzburg \& Syrovatskii 1964).
If indeed the ionization state evolves during the
Pop II epoch, the effective change in $\Lambda$ can affect the cosmic ray
spectrum independent of the considerations
in eqns.\ (\ref{eq:lameff}) and (\ref{eq:lamoft}).

\subsection{Spectrum}
\label{sect:spec}

Both WSSOF and PCV have independently argued for a flatter spectrum in early
cosmic rays than that presently observed.  The key observation supporting
this argument is the B/Be ratio.
Although the observational
uncertainty here is clearly
large, the
current preliminary numbers are
low compared to
the present day observed value of $15 \pm 3$ (with e.g.\ Mathews,
Walker, and Viola (1985) predicting $\simeq 17$ for Pop I),
and to
the Pop II value of $\simeq 14$ calculated in WS (using a transported
spectrum $\phi(T) \propto (T+m_p)^{-2.7}$ and Pop I C:N:O ratios).

The argument for a flat early cosmic ray spectrum
arises if one takes a low central value for the Pop II B/Be ratio
seriously.  (Remember values below 7 are impossible in any spallation
model, and all hypothetical stellar processes suggested thus far
act to increase B/Be.)
Be and B arise from the same spallation processes, so
there are a limited number of available differences to exploit
in fixing the ratio.  Their differences in mass and charge lead to
different ionization losses and ranges, and consequently different
confinement for the two elements (at a given energy per nucleon).
However, with the ionization
ranges scaling law of eq.\ (\ref{eq:irscaling}), it is clear that
$R_{B}(T) < R_{Be}(T)$, and thus energy loss and thermalization
effects (c.f.\ eq.\ \ref{eq:sfac})
favor B rather than Be and so act to {\em increase} the B/Be ratio.
The other important difference between Be and B production rates
lies in their spallation cross sections.
As argued in section \ref{sect:conf}, the differences make the
B/Be ratio sensitive to the cosmic ray spectrum, with a lower
B/Be for flatter spectra.

Another argument for a flatter spectrum has been offered by PCV
(but weakened by Malaney and Butler's (1993)
arguments; see Prantzos \& Cass\'{e} (1993)).
PCV note that the low observed Li/Be ratio (once the primordial Li
component has been subtracted)
is more easily reproduced by a
flat spectrum because of exponential dropoff of $\alpha+\alpha$ cross
sections.

In addition to the effect of changes in the confinement, the propagated
spectrum obviously depends on the source spectrum.
PCV choose as a source spectrum
a power law in momentum, $q(p) \propto p^{-2}$, a type of spectrum
appearing in models of cosmic ray shock acceleration.  Such models
assume acceleration via scattering off of supernova shock waves, and
have been successful in
reproducing galactic and heliospheric spectra in detail
(c.f.\ the  Blanford
\& Eichler (1987) review).
Note that the PCV spectral index of 2
is the universal limiting value for very large Mach number shocks.
We note, however,
that despite the success of and active interest in this model,
there remains disagreement upon the the appropriate model for acceleration.
It is thus worthwhile to examine the sensitivity of the LiBeB yields
to the assumed spectrum, where
there remains some flexibility to choose the spectral index.

Most important for LiBeB production is the choice of spectral type.
While shock acceleration models suggest source power laws in
magnetic rigidity and hence momentum are good candidates, the measured
(and hence propagated) cosmic ray flux at earth is also consistent
with a source spectrum in total energy per nucleon.
The two are of
course difficult to distinguish at the relativistic energies at which
the cosmic ray measurements are unaffected by solar modulation effects.
Furthermore, solar effects introduce large uncertainties at lower energies,
rendering unreliable the data that could make this distinction.
In addition, a power law in total energy has the advantage
that it remains finite at low energies, thus providing a mockup of
physical processes responsible for preventing a divergence in the source
at low energies.

Thus we have tried source spectra of power laws both
in momentum and in total energy per nucleon, i.e.
\begin{eqnarray}
\nonumber
q(p) & \propto & p^{-\gamma}  \\
\label{eq:psource}
& \Longrightarrow & q(T) \propto
               (T + m_p)(2m_pT + T^2)^{-(\gamma+1)/2} \; , \; {\rm and} \\
\label{eq:esource}
q(T) & \propto & (T + m_p)^{-\gamma}
\end{eqnarray}
with $m_p$ the proton mass.
For the shock acceleration model--inspired momentum spectrum,
the index $\gamma$ is given by
\beq
\gamma \approx  \left( 2 + \frac{4}{M^2} \right)
\eeq
for the case of large Mach number
$M = v_{shock}/c_{sound}$, the ratio of shock velocity
to the sound speed in the shocked gas.
Note that the Mach number depends upon
the local ISM sound speed, which in turn depends upon, e.g. the ISM density
and pressure.
If the early galaxy, or its primordial protogalactic building blocks,
were to be significantly hotter or denser than
today, then potentially the cosmic ray spectrum could be quite different.

For the total
energy spectrum we take $\gamma = 2.7$ , which is consistent with the
measured, transported
cosmic ray spectrum.
Note however, that this choice is
inappropriate for the simple leaky box with
energy dependent escape length.  In this case, we live in the
confinement region (the galaxy), and so the escape energy dependence
is folded into that of the source.
The proper source index is thus
$\gamma_{source} = \gamma_{obs} - \beta \approx 2.1$, with $\beta$ the
high-energy
index of the energy-dependent
$\Lambda$ from eq.\ (\ref{eq:gmpld}).
This is not the case in, for example,
the nested leaky box model, in which we do not live in the confinement
region (which shrouds the sources),
and so we do not see its energy dependence:
$\gamma_{source}=\gamma_{obs}$.  Again, the model uncertainty argues
for examination of a range of source indices.

Note the
dramatic effect that the spectral index has
on the Li/Be and B/Be plots (figs.\ \ref{fig:lb-lam}
and \ref{fig:bb-lam}).  The large low energy flux of
the momentum spectrum makes this spectrum softer than one in total
energy, and the resulting elemental ratios behave as one would expect
for such a soft spectrum.  As discussed above in section \ref{sect:chemev},
low energy resonance behavior is emphasized and leads to large Li/Be
and B/Be ratios.

We have allowed the spectral index $\gamma$ to vary
over the range $2 \le \gamma \le 3$.  Results appear in figures
\ref{fig:lb-ind} and \ref{fig:bb-ind} from which it is
apparent that the LiBeB production rates
from a source with a momentum spectrum
are more sensitive to changes
in spectral index than are the rates for a source total energy law.
Indeed it is clear from fig.\ \ref{fig:lb-ind} that for source spectra in
momentum, indices with $\gamma \ga 2.5$ lead to Li-to-Be relative production
rates in ratios approaching $10^3$.  This is clearly undesirable
given that all observed Be abundances are $\ga 10^{-13}$, thus implying
associated Li production larger than the observed plateau, even without
the large primordial Li component.

While the relative B and Be rates (figure
\ref{fig:bb-ind}) do not vary as dramatically for the momentum spectrum,
here too they are much more sensitive than  the same ratios given a total
energy spectrum.  Additionally, the same range of indices that overproduce
Li relative to Be give the
largest values  for  B relative to Be
as compared to
low values for the current preliminary but uncertain data.
Nevertheless, it is intriguing that the range of indices that fail for the
Li to Be ratio seem to also do the worst for B to Be.
Thus momentum spectra with indices steeper than about 2.5 are less favored
at the present time.
Note that we get no such constraints for source spectra in total
energy, as we see only mild sensitivity to a source index
for Li to Be (for which all values are completely acceptable), and
completely insensitivity for B to Be.

In figure \ref{fig:lb-lamind} we plot
\dlibe\ vs $\Lambda$, as in figure \ref{fig:lb-lam}, but
with different spectral indices.  For each spectral type,
the production ratio grows with the steepness of the spectrum,
the growth being more pronounced for the scale-free momentum law.
Figure \ref{fig:bb-lamind} is a similar plot for the \dbbe\ ratio.  Note
that this ratio is mildly sensitive to the momentum spectral index,
which can change \dbbe\ by 1 to 2 units, i.e.\ 10-20\%.  Such a change
might be detectable as observations improve.  The \dbbe\ ratio is even
less sensitive to the index of a source with a total energy spectrum.
The change over the whole range is of order 5\%.

As we have alluded to earlier, a direct comparison of the data and the model
calculations given here is extremely difficult but can be suggestive.
The ratio of of $Li/Be$ given in table 1 contains the primordial
Li component as well as the cosmic-ray produced Li.
In some cases this correction might be quite large (a factor of 10 or more)
so that the ratios in the table represent extreme upper limits.
If we had a specific cosmic-ray nucleosynthesis model, one could, as shown in
Olive and Schramm (1992) extract a Big Bang Li abundance.  For the
cosmic-ray nucleosynthesis model used in WSSOF this led to a value
$[{\rm Li}]_{BB} = 2.01 \pm 0.07$, however as we can see from figures
\ref{fig:lb-fe}{\it a} and \ref{fig:lb-fe}{\it b}, there
is significant model dependence in the Li/Be ratio.
Also, the absolute [Li] values are plagued by the poorly given
systematic errors mentioned in section \ref{sect:data}.
Furthermore, as we have also noted, the element ratio will also be affected
by chemical evolution.
Thus one can with certainty only require the data points
as given in the table to lie above any model curve in the figures.
The trend in the data is also seen in the figure \ref{fig:lb-fe},
i.e. at high metallicities
the Li/Be ratio is smaller.  Note that a large primordial subtraction
would favor the lower ratios found with the cosmic-ray energy source
spectrum.

To get a feeling for the comparison between data and these models, we can
make a primordial subtraction corresponding
to $[{\rm Li}]_{BB} = 2.0$.  It hard
to make the primordial \li7 abundance much smaller (Walker et al.~1993;
Krauss \& Romanelli, 1990)
and if it were much larger there would be no room for
cosmic-ray produced Li.  In Figure \ref{fig:libe-fedat}
we show the values of $Li/Be$ with
the above primordial subtraction. Error bars are too large to display.
We also plot the ratio from the energy
and momentum-spectra for $\Lambda = 100\grams$ as a function of [Fe].

\subsection{Other Model Features}
\label{sect:omod}

The conventions for reporting spallation/fusion cross sectional data
(e.g.~Read \& Viola 1984)
are strictly speaking incomplete for calculations of LiBeB production,
and thus require a model to make some assumption regarding the kinematics
of these interactions.
Specifically, the tabulated data and semi-empirical fits give the
{\em total} cross section for LiBeB production at a given incident
energy.  As seen explicitly in eq.\ (\ref{eq:sourceterm}), however,
we require the full, {\em differential} cross section
$d \sigma_{ij}^A/d T_{LiBeB} \left(T_{LiBeB},T_{CR}\right)$.
Meneguzzi et al.\ (1971) and many subsequent authors
address this issue by assuming that the differential cross section
is very sharply peaked around those values of the LiBeB energy
for which $T_{LiBeB} = T_{CR}$.  They put
\beq
\label{eq:menkin}
\frac {d \sigma_{ij}^A}{d T_{LiBeB}} \left(T_{LiBeB},T_{CR}\right)
= \sigma_{ij}^A(T_{CR}) \; \delta\left(T_{LiBeB}-T_{CR}\right) \,\, .
\eeq
with $\sigma_{ij}^A(T_{CR})$ the tabulated cross section.

While the approximation of
sharp peaking over a small range of product energies appears to be
borne out experimentally in the cases where it has been checked, the
other approximation, that of equal energies (per nucleon) in the initial
and final state is only valid for the case of heavy cosmic rays
on light targets.  We have relaxed the latter assumption, allowing
for the CR energy to be related to the LiBeB
energy via some function $E_f(T_{CR})$, changing eq.\
(\ref{eq:menkin}) to
\beq
\frac {d \sigma_{ij}^A}{d T_{LiBeB}} \left(T_{LiBeB},T_{CR}\right)
= \sigma_{ij}^A(T_{CR}) \; \delta\left(T_{LiBeB}-E_f(T_{CR})\right) \,\, .
\eeq

With this more general formalism we have investigated two possible
kinematic behaviors.  One
ansatz is that in the center of momentum frame the LiBeB is produced at rest,
with the lighter debris moving in such a way as to conserve energy.
This is the assumption we have made throughout this paper.
One should compare this to
another common assumption that
in the frame in which the light target is at rest,
the kinetic energy per nucleon of the daughter nucleus is the same
as that of the heavy CR parent.  We have plotted the $\dlibe$ ratio
for these two cases in figure \ref{fig:libe-lamaltkin}.

The issue of spallation kinematics is raised if one considers
an alternative to the usual production scenario we have presented so far.
Duncan et al.\ (1992) and PCV both attempt to address the question of
the Be vs Fe slope with the suggestion that perhaps most of the
LiBeB production is not from light (p, $\alpha$) cosmic rays on
heavy (CNO) ISM nuclei, but the reverse:  heavy cosmic rays on light ISM
nuclei.  Furthermore,
if the heavy cosmic rays do not have the ISM abundance
or something near to it,
then this could yield significantly different Be and B
production.  Increased cosmic ray heavies would result, for example, if
supernovae were to accelerate their own ejecta.
The evidence is not definitive
on this issue, but there are possible problems in
maintaining consistency with an assumed
first ionization potential
dependence of the cosmic ray source abundances.

At any rate, we have investigated the dependence of the spallation yields
on forward and reverse kinematics.  Figure \ref{fig:libe-lamrkin}
compares the $\dlibe$ ratio for the cases of normal kinematics (i.e.\
LiBeB synthesis primarily from light CR on heavy ISM with a small
admixture of the reverse) to reverse kinematics, in which we assume
$y_{CNO}^{CR} \approx y_{CNO}^{solar}$, and thus dominate the LiBeB
production.  Note that here the $\dlibe$ ratio is much smaller, as
the $\alpha+\alpha$ process is not as dominant as in the usual early
galactic case.

Note that, in considering the thermalization of LiBeB,
authors from Meneguzzi et al.~(1971) to PCV apparently use
$S \equiv 1$ over all energies in the case of light cosmic
rays on heavy targets.  These authors note that the daughter nuclei
move more slowly than the cosmic ray parent and so are likely to be
stopped.  However, as we are considering cases where the flux can
be very hard, then we may still have quite energetic daughter
nuclei and it is important to allow properly
for incomplete stopping.
Note that Li, which is made primarily
by $\alpha+\alpha$ fusion, is thus the only spallogenic
element we consider for which (in the dominant production process)
the cosmic ray projectile not much lighter than the ISM target.
Consequently, the Li will be the fastest daughter nuclei,
and thus accurate treatment of trapping is particularly important.
To compare with previous work, we have
computed yields with $S \equiv 1$ for both kinematics.
Results appear in figure \ref{fig:libe-lams1}.

We see in figure \ref{fig:libe-lams1} that the direction
of the effect depends upon the spectral type.  For a momentum source,
one gets a higher \dlibe\ ratio when allowing for realistic trapping
of LiBeB.  The change, ranging from about 40\% at large $\Lambda$
down to 10\% at small ones, arises because
the Li-to-Be production ratio is the
smallest in the energy regimes; however reactions
at these energies produce fast products which
are less likely to be thermalized than slow ones.
Put differently, if $T_{hi} \gg T_{low}$ are different LiBeB
energies, then
in the language of eq.\ \ref{eq:sfac}, $S(T_{hi}) \ll S(T_{low})$
This is not the whole story for a source law in energy.
Here, the spectral peak
is much more pronounced
and at a high energy ($\sim m_p$)
than the (lower energy) peak of the flux from the momentum source.
The flux from the energy source thus {\em rises} steeply  at
the ($\sim 100 $MeV) dropoff in the
Li production cross sections.
Consequently, the majority of Li production occurs in
the beginning of the high energy tail of the cross section.
By disfavoring the high energy products through incomplete stopping,
one removes the faster Li made at the highest overlap of cross section
and flux.  With $S=1$, however, these fast daughters are included
and so raise the Li production relative to Be.

Let us now look at the dependence of the \dlibe\ production rate
on the asymptotic value of cross section for
$\alpha + \alpha \rightarrow \li{6,7}$.  This is a fusion process
and is physically different from the other spallation processes.
The data show that these cross sections decrease exponentially with
energy,
although the highest energy at which a definite cross
section has been reported is only about 50 MeV/nucleon.
Evidence for the continued dropoff comes instead from
upper bounds to the cross sections, which go out to $\sim 200$ MeV/nucleon,
with no plateau detected.
In their cross section tabulation, Read and Viola (1984)
point out the uncertainty in the high-energy $\alpha+\alpha$ reaction,
and recommend adopting an exponential decrease with energy, though
they do allow for the possibility
that a plateau might nevertheless
eventually be found to
exist.
We feel that
an exponential dropoff is the more appropriate
description of a fusion reaction, but some work (notably PCV) has
employed such a plateau.
We have therefore examined the \dlibe\ ratio with and without
such a plateau, and find that it can have a significant effect, particularly
for cases of large confinement or spectral index near 2, as these
emphasize the high energy events in which the differences are important.
Figure \ref{fig:spallxs} compares the \dlibe\ ratio with and without
the cross section plateau.  While the differences are not large
for a momentum source ($\la 10\%$), for a source power law in total energy,
the effect is about a factor of 2.  Since we see no evidence as yet
for a plateau, in
all our other computations none is assumed to exist.
However, as the effect of such a plateau is significant,
we feel it is very important for experiments to investigate the
behavior of the $\alpha+\alpha \rightarrow \li{6,7}$ at high energy.

Finally, we note the important but not decisive
contribution
to the overall uncertainty
of experimental uncertainties in cross sections and
in energy losses.  Cross section uncertainties
are typically at a level of $\delta \sigma/\sigma \simeq 10 - 20\%$.
While these errors are large enough to be significant, they
clearly are smaller than the other model effects we have noted above,
consequently making the model choices important ones.
Energy losses are better known, with
$\delta b/b \simeq \delta R/R \simeq $ 5\% at low energies
(where they are unimportant), and $\la 2\%$ at higher energies.
These errors do not significantly contribute to our overall uncertainty.

\section{Gamma Ray Production}
\label{sect:gammaray}

The same passage of cosmic rays through the ISM that
produces LiBeB must also create $\gamma$-rays,
as noted by
Silk and Schramm (1993), Fields, Schramm and Truran (1993),
and Prantzos and Cass\'e (1993).
The primary $\gamma$-ray source from the nuclear cosmic ray
component\footnote{As opposed to the cosmic ray electrons,
which produce lower energy $\gamma$-rays by brehmsstrahlung.  We do not
explicitly consider the electron component here, but
Prantzos and Cass\'e argue that it should scale with the nuclear component
as it does today, and using this scaling they calculate the
contribution to the isotropic $\gamma$-ray background.}
arises from inelastic
scattering of cosmic rays on the ISM,
with $p + p \rightarrow \pi^0 \rightarrow 2\gamma$.
Thus the $\gamma$-ray spectral peak is around $m_\pi/2$.

The total (isotropic) $\gamma$-ray production rate
by the nuclear component of the
cosmic rays, per second per cubic centimeter, is given by
\beq
\Gamma = n_H \int_{T_{th}}^\infty \, dT
  	\phi_p(T) \sigma_{pp}^\gamma (T)
\eeq
in an obvious notation.  Note that the
calculation is similar to that for LiBeB production.
Indeed this process occurs under very similar conditions to
Li production, as both processes derive from reactions between
protons and $\alpha$ nuclei, which in the early galaxy are undepleted
both in the cosmic rays and in
the ISM.  Thus we may expect to scale the $\gamma$-ray production
to that of Li via $\sigma(\alpha+\alpha \rightarrow$ Li) to the
$\sigma(p+p \rightarrow \gamma)$
cross section ratio.  One should note, however, that such a simple scaling
ignores the different energy dependences of these processes, and
ignores the issue of the incomplete trapping of the Li fusion products.

Because $\gamma$-ray production so closely follows that of Li, the model
dependences will be similar.
In particular, the effects of cosmic ray evolution
will be similar, as will the uncertainties in scaling the $\gamma$-ray
production to the Be.
In a subsequent paper on specific galactic evolution
models we intend to quantify the $\gamma$-fluxes and use them
as a complementary probe to the LiBeB abundance
behaviors described here.

\section{Conclusions}
\label{sect:con}

In this paper we showed in detail that there are many
uncertainties about early cosmic ray behavior
and therefore a range of possible assumptions
{}from which one must choose in building a
model for this behavior.
Furthermore, the variance in possible early cosmic ray models can
have a significant effect of
on the calculation of cosmic ray production of LiBeB (and $\gamma$-rays)
in the
early galaxy.  The variance in the models derives both from
uncertainties about present cosmic rays and from uncertainties about
the evolution of their injection and acceleration, as well as confinement.

Because these uncertainties are difficult to resolve, we conclude that
future work in modeling early cosmic rays must
allow for cosmic ray
model variances and evolution.
As shown in section \ref{sect:spec}, the choice of source spectral type
is significant yet not settled even for today's cosmic rays,
with power laws both in total energy per nucleon and in momentum
being allowed.
Also, we show that cosmic ray evolution derives not only from
changes in source strength, but also from changes in confinement,
as detailed in sections \ref{sect:chemev} and \ref{sect:conf}.
Evolution of confinement can have important effects, for example in
altering the na\"{\i}vely expected Be-to-Fe slope, as outlined here
and shown in the detailed model of PCV.  Additional considerations
are noted in section \ref{sect:omod}, most notably the effects of inelastic
nuclear collisions on confinement, as have been noted by
Malaney and Butler (1993).

In light of our findings we now turn to the work to date on
early cosmic ray production of LiBeB, namely the relatively model-independent
work of WSSOF and its sequels, and the detailed cosmic ray and chemical
evolution model of PCV.
The flatness of the
curves in figure \ref{fig:bb-lamind} justifies the claim in WSSOF that
the B/Be ratio in cosmic ray models is indeed insensitive to the details
of cosmic ray models.  In addition, because Be and B are pure spallation
(as opposed to $\alpha + \alpha$ fusion) products, their production ratio
\dbbe\ is independent of [Fe/H] and varies only with the C:N:O ratios of
targets.  To the extent that this ratio remains fairly constant, we can
conclude that the B/Be ratio is insensitive to chemical evolution as well.
We consequently find B/Be to be well-chosen as a model-independent
signature of the spallation process.

The Li isotopic ratio, discussed in SFOSW,
\li6/\li7 shows a similar independence to cosmic ray model features
but does have a moderate sensitivity to
chemical evolution via the ISM ratio He/CNO, which
sets the relative contributions of the $\alpha+\alpha$ fusion
versus spallation
contributions to \li6/\li7.  We find that $d\li7/d\li6$ is always
within $\sim 30$\% of unity
for all spectra and escape parameters considered here.
This uncertainty is much smaller than that of the \li6/\li7 ratio
of Smith et al.~(1992).  Consequently we find SFOSW's assumption
of a temporally  constant $d\li6/d\li7$ production ratio to be appropriate.

However, we find that the ``zeroth order model'' of WS, WSSOF, and SFOSW to
be overly simplistic
not only in its lack of chemical evolution, but also in
its lack of any kind of cosmic ray evolution.  The effect on predictions of
this model is unimportant for the B/Be limits, but
their calculations of the
Li/Be ratio are, as we have noted, only a lower bound to the actual
ratios.

The PCV model does allow for chemical as well as cosmic ray evolution, and
in particular they allow for evolution of cosmic ray confinement.
We thus find their model indeed to be a useful examination of LiBeB
production.  Our caution is that their model is a specific one made
with particular, albeit reasonable, assumptions.
Their specific cosmic ray source model we find to be
well-motivated and reasonable,
but not a unique choice and we would urge that future work consider different
momentum source indices, as well as source spectra in total energy.
Furthermore, PCV admit that
their particular implementation is to some degree arbitrary.  In the first
place, it is not obvious which parameters control $\Lambda$.
Secondly,
even given a particular scaling prescription, it is not
trivial to model the input parameters (e.g.\ galactic
scale height and magnetic field) accurately.  PCV assume
$\Lambda$ depends upon the scale height of the collapsing disk, but they
do not compute the collapse explicitly, and so impose an
{\em ad hoc} time dependence.  Clearly a firm model for $\Lambda$
evolution is crucial, but at present is unavailable.  We agree with
PCV that more theoretical work in this area is needed.

In summary, we feel that improved data on PopII LiBeB abundances
with consistent controls on systematics might lead to
an ability to constrain models for the origin and
propagation of cosmic rays in the early galaxy.

\vskip 1.5cm
\noindent  We thank Doug Duncan, Gary Steigman, Simon Swordy, Mike Thayer,
Julie Thorburn, and Terry Walker
for useful discussions.  This work was supported in part by the
DOE (at Chicago and Fermilab), by NASA and NSF at Chicago,
 and by NASA through NAGW-2381 (at Fermilab). This work was also
supported in part by  DOE grant DE-AC02-93ER-40105 at Minnesota.
The work of KAO was in addition supported by a Presidential Young
Investigator Award.

\vskip 2 cm

\beginapjbib

\bibitem Asakimori, K., et al.~1993, in Proc.~of the 23rd Int.\ Cosmic
Ray Conf.\ (Calgary), 2, 25

\bibitem Blanford, R., and Eichler, D. 1987, Phys Reports, 154, 1

\bibitem Boesgaard, A.M., \& King, J. 1993, preprint

\bibitem Bystricky, J., Lechanoine-Leluc, C., \& Lehar, F. 1987,
J. Physique, 48, 199

\bibitem Buckley, J., Dwyer, J., M\"uller, D., Swordy, S., and Tang, K.K.
1993, ApJ submitted

\bibitem Ceasarsky, C.J. 1980, ARAA, 18, 289

\bibitem Ceasarsky, C.J. 1987, in Proc.\ of the 20th Int.\
Cosmic Ray Conf.\ (Moscow), 8, 87

\bibitem Dearborn, D.S.P., Schramm, D.N., Steigman, G., \&
Truran, J. 1988 ApJ, 347, 455

\bibitem Duncan, D.K., Lambert, D.L., \& Lemke, M. 1992, ApJ, 401, 584

\bibitem Feltzing, S., \& Gustaffson, B. 1994, ApJ, 423, 68

\bibitem Fields, B. D., Olive, K.A., \& Schramm, D. N. 1994, A\&A, submitted

\bibitem Fields, B. D., Schramm, D. N., \& Truran, J. W. 1993 ApJ, 506, 559.

\bibitem Garcia-Mu\~{n}oz, M., Simpson, J.A., Guzik, T.G., Wefel, J.P.,
\& Margolis, S.H. 1987 ApJS, 64, 269

\bibitem Gilmore, G., Edvardsson, B., \& Nissen, P.E. 1992a, AJ, 378, 17

\bibitem Gilmore, G., Gustafsson, B., Edvardsson, B., \& Nissen, P.E. 1992b,
Nature, 357, 379

\bibitem Ginzburg, V.L., \& Syrovatskii, S.I. 1964 The Origin of Cosmic Rays,
(Pergamon: New York), 121

\bibitem Hobbs, L.M., \& Duncan, D., 1987, AJ, 317, 796

\bibitem Hobbs, L.M., \& Thorburn, J.A., 1991, ApJ, 375, 116

\bibitem Janni, J.F. 1982, Atomic Data \& Nuc Data Tables, 27, 34

\bibitem Krauss, L.M., \& Romanelli, P. 1990 ApJ, 358, 47

\bibitem Malaney, R.A., \& Butler, M.N. 1993, ApJ, 407, L73

\bibitem Meneguzzi, Audouze, J., \& Reeves, H. 1971 A\&A, 15, 337

\bibitem Meyer, J.P. 1972 A\&ASup, 7, 417

\bibitem Molaro, P., Castelli, F., \& Pasquini, L. 1993, Origin and Evolution
of the Light Elements, ed.\ N. Prantzos, E. Vangioni-Flam, \& M. Cass\'e
(Cambridge: Cambridge University Press), 153

\bibitem Northcliffe, L.C., \& Schilling, R.F. 1970, Nuc Data Tables, A7, 233

\bibitem Olive, K. A., \& Schramm, D. N. 1993 Nature, 360, 439

\bibitem Olive, K.A., Prantzos, N., Scully, S., \& Vangioni-Flam, E. 1993,
ApJ, in press

\bibitem Pilachowski, C.A., Sneden, C., \& Booth, J. 1993, ApJ, 407, 699

\bibitem Prantzos, N. 1992, in Nuclei in the Cosmos, ed.\ F. Kappeler
\& K. Wisshak (Philadelphia:IOP Publishing), 471

\bibitem Prantzos, N., \& Cass\'{e}, M. 1993, Saclay preprint

\bibitem Prantzos, N., Cass\'{e}, M., Vangioni-Flam, E.
1993 ApJ, 403, 630 (PCV)

\bibitem Read, S.M., \& Viola, V.E., 1984

\bibitem Rebolo, R., Molaro, P., Abia, C., \& Beckman, J.E.
1988a A\&A, 193, 193

\bibitem Rebolo, R., Molaro, P., \& Beckman, J. 1988b, A\&A, 192, 192

\bibitem Reeves, H., Fowler, W. A., \& Hoyle, F. 1970 Nature, 226, 727

\bibitem Ryan, S., Bessel, M., Sutherland, R., \& Norris, J.
1990, ApJ, 348, L57

\bibitem Ryan, S., Norris, J., Bessel, M., \& Deliyannis, C.
1992 ApJ, 388, 184

\bibitem Silk, J., \& Schramm, D.N., 1993, ApJ, 393, L9

\bibitem Smith, V.V., Lambert, D.L., \& Nissen, P.E. 1992, ApJ, 408, 262

\bibitem Spite, F., Maillard, J.P., \& Spite, M. 1984, A\&A, 141, 56

\bibitem Spite, F., \& Spite, M. 1982, A\&A, 115, 357

\bibitem \rule [0.33em]{1.5cm}{0.1mm} .\ 1986, A\&A, 163, 140

\bibitem Steigman, G., Fields, B. D., Olive, K. A., Schramm, D. N.,
\& Walker, T. P. 1993 ApJ, 415, L35 (SFOSW)

\bibitem Steigman, G., \& Walker, T. P. 1992 ApJ, 385, L13 (SW)

\bibitem Thomas, D., Schramm, D.N., Olive, K.A., \& Fields, B.D.
1993, ApJ, 406, 569

\bibitem Townsend, L.W., \& Wilson, J.W. 1985,
Tables of Nuclear Cross Sections
for Galactic Cosmic Rays, NASA Ref Pub 1134

\bibitem Vangioni-Flam et al.~1990, ApJ, 364, 568

\bibitem Walker, T. P., Mathews, G. J., \& Viola, V. E. 1985 ApJ,
299, 745

\bibitem Walker, T. P., Steigman, G., Schramm, D. N., Olive, K. A.,
\& Fields, B. D. 1993 ApJ, 413, 562 (WSSOF)

\bibitem Walker, T. P., Steigman, G., Schramm, D.N., Olive, K.A., \&
Kang, H. 1991, ApJ, 376, 51

\bibitem Woosley, S.E., Hartmann, D.H., Hoffman, R.D., \& Haxton, W.C. 1990,
ApJ, 356, 272

\endapjbib

\hfill \pagebreak

\vskip 2cm
\begin{center}
{\bf FIGURE CAPTIONS}
\end{center}

\newcounter{bean}
\begin{list}
{\bf Figure \arabic{bean}:}{\usecounter{bean}
                            \listparindent=1.1cm}

\item \label{fig:fluxnorm}
({\it a})
The cosmic ray proton flux and $\alpha$ spectra $\phi$ compared to source
spectra ($q(p) \propto p^{-2}$) scaled as $\Lambda q$.  Note that
in the limit of no energy losses (a good approximation at high energy)
we have $\phi = \Lambda q$, and so departure from this scaling indicates
the effect of energy losses.  Note also that the $\alpha$ flux scales
very well with the proton flux
as the He cosmic ray abundance 0.08.
That this proportionality is accurate over the whole energy range
reflects the
similar energy loss behavior due to the scaling $A/Z^2$. \par
({\it b}) As in ({\it a}), for $q(T) \propto (T+m_p)^{-2.7}$.

\item \label{fig:fluxlam}
({\it a}) The transported cosmic ray proton flux for
a range of
confinement parameters:
$\Lambda = 10$, 30, 100, 300, and 1000 \grams.
The source is a power law in momentum, $q(p) \propto p^{-2}$. \par
({\it b}) As in ({\it a}), for a source power law in total energy,
$q(T) \propto (T+m_p)^{-2.7}$.

\item \label{fig:lb-lam}
The Li to Be production ratio as a function of pathlength.
The solid line is for a source spectrum that is a
power law in momentum, $q(p) \propto p^{-2}$; the
dashed line is for a source power law in total
energy, $q(T) \propto (T+m_p)^{-2.7}$.
We put [C/H] = [N/H] = -2.5,
[O/H] = -2.0.

\item \label{fig:bb-lam}
The B to Be ratio as a function of pathlength, plotted
as in fig.\ \ref{fig:lb-lam}.

\item \label{fig:lb-fe}
({\it a}) The production ratio of Li to Be as a function of
metallicity [Fe/H],
for a source $q(p) \propto p^{-2}$, and for a range of conement parameters:
$\Lambda$ = 10, 30, 100, 300, and 1000 \grams.
See discussion in the text regarding
comparison to observations. \par
({\it b}) As in ({\it a}), for a source $q(T) \propto (T+m_p)^{-2.7}$.

\item \label{fig:gmpld}
A plot of the pathlength distribution of equation (\ref{eq:gmpld}),
chosen in Garcia-Mu\~{n}oz et al.\ (1987) to provide the best fit to
the observed B/C ratio.

\item \label{fig:lcomplibe}
({\it a}) The \dlibe\ ratio for $\Lambda(T) = \Lambda_0$ (solid line), and
for $\Lambda(T)$ given by equation (\ref{eq:gmpld}), with $\Lambda_0$ as
indicated on the figure (broken line).  The source spectrum is
$q(p) \propto p^{-2}$, and ISM abundances are [C/H] = [N/H] = -2.5
and [O/H] = -2.0.  \par
({\it b}) As in ({\it a}), for a source spectrum
$q(T) \propto (T+m_p)^{-2.7}$.

\item \label{fig:lcompbbe}
({\it a}) The \dbbe\ ratio for $\Lambda(T) = \Lambda_0$ (solid line), and
for $\Lambda(T)$ given by equation (\ref{eq:gmpld}), with $\Lambda_0$ as
indicated on the figure (broken line).  The source spectrum is
$q(p) \propto p^{-2}$ and ISM abundances are [C/H] = [N/H] = -2.5
and [O/H] = -2.0.  \par
({\it b}) As in ({\it a}), for a source spectrum
$q(T) \propto (T+m_p)^{-2.7}$.

\item \label{fig:lamwinel}
({\it a}) The \dlibe\ ratio for $\Lambda = \Lambda_{esc}$, the usual case
(solid line), and for
$\Lambda(T)$ given by eq.\ (\ref{eq:lameff}), (broken line), where we have
plotted against $\Lambda_{esc}$.  The source spectrum is
$q(p) \propto p^{-2}$ and ISM abundances are [C/H] = [N/H] = -2.5
and [O/H] = -2.0.  \par
({\it b}) As in ({\it a}), for a source spectrum
$q(T) \propto (T+m_p)^{-2.7}$.

\item \label{fig:lb-ind}
The ratios of Li to Be production rates, shown
as a function of source spectral index $\gamma$ for
a source spectrum in momentum and in total energy.
Note that the yields from the momentum spectrum are very sensitive to the
index adopted, and the ratio of rates can become unacceptably high
in this case.  In contrast, the yields from the total energy spectrum
are relatively insensitive to spectral index, and are all at acceptably
low levels.

\item \label{fig:bb-ind}
As in fig.\ \ref{fig:lb-ind} for the ratio of B to Be rates; the same
trends are seen.  Note however that the B/Be ratio is negligibly sensitive to
the total energy index, and is only mildly sensitive to that for the
momentum law.

\item \label{fig:lb-lamind}
({\it a}) Curves of \dlibe, as a function of $\Lambda$,
for different spectral indices $\gamma$.  The source is
$q(p) \propto p^{-2}$. \par
({\it b}) As in ({\it a}), for
$q(T) \propto (T+m_p)^{-2.7}$.

\item \label{fig:bb-lamind}
As in fig.\ \ref{fig:lb-lamind}, for the \dbbe\ ratio.

\item \label{fig:libe-fedat}
Similar to figure \ref{fig:lb-fe}.  We plot \dlibe\ for each source type
at $\Lambda=100\grams$.  The data, as described in the text, has been
corrected assuming the minimal allowed primordial \li7 production,
$[\li7]_{BB}$ = 2.0 and thus (linearly) subtracting this amount from the Li
vales of table 1.  See discussion in the text regarding caveats in comparing
the theoretical curves, which are upper bounds, to the data.

\item \label{fig:libe-lamaltkin}
The $\dlibe$ ratio for different determinations of the daughter nucleus
kinematics.  The solid curve is calculated assuming that the daughter nucleus
is stationary in the center of momentum frame of the parents; this
is the standard used throughout this paper.  The dashed curve is calculated
assuming that, in the rest frame of the light parent, the daughter
kinetic energy per nucleon is equal to that of the heavy parent.
The source spectrum is $q(p) \propto p^{-2}$ for curves ({\it a} and ({\it b},
and $q(T) \propto (T+m_p)^{-2.7}$ for curves ({\it c} and ({\it d}.
Additionally, we put [C/H]=[N/H]=-2.5, [O/H]=-2.0.

\item \label{fig:libe-lamrkin}
({\it a})
The $\dlibe$ ratio for (solid line)
cosmic ray CNO abundances equal to ISM abundances
($y_{CNO}^{CR} = y_{CNO}^{ISM}$, the usual case), and for (dashed line)
cosmic ray CNO abundances fixed to be solar
($y_{CNO}^{CR} \gg y_{CNO}^{ISM}$).  The source spectrum is
$q(p) \propto p^{-2}$ and ISM abundances are [C/H] = [N/H] = -2.5,
and [O/H] = -2.0 \par
({\it b}) As in ({\it a}), for a source spectrum
$q(T) \propto (T+m_p)^{-2.7}$.

\item \label{fig:libe-lams1}
The $\dlibe$ ratio with incomplete LiBeB trapping, i.e.\ with
$S_{\rm LiBeB} = \exp(R/\Lambda)$ (solid line), and with complete
LiBeB trapping ($S_{\rm LiBeB} = 1$).
The source spectra are as labeled, and we put
and [C/H]=[N/H]=-2.5, [O/H]=-2.0.

\item \label{fig:spallxs}
({\it a}) The \dlibe\ ratio in the absence (solid line) and presence (broken
line) of a high energy plateau in the $\alpha+\alpha \rightarrow \li{6,7}$
cross section.  The source spectrum is
$q(p) \propto p^{-2}$ and ISM abundances are [C/H] = [N/H] = -2.5
and [O/H] = -2.0.  Note the small ($\sim 10\%$)
increase with the plateau. \par
({\it b}) As in ({\it a}), for a source spectrum
$q(T) \propto (T+m_p)^{-2.7}$.  Here the difference is larger (a
factor of $\sim 2$),
as the flux is harder than that from the momentum flux, and so the high
energy cross section behavior is more important.

\end{list}

\end{document}